\documentclass[journal]{IEEEtran}
\usepackage{amsmath,amsfonts}
\usepackage{tabularx}
\usepackage{algorithmic}
\usepackage{array}
\usepackage[caption=false,font=normalsize,labelfont=sf,textfont=sf]{subfig}
\usepackage[style=ieee,sorting=none]{biblatex}

\usepackage{textcomp}
\usepackage{stfloats}
\usepackage{url}
\usepackage{verbatim}
\usepackage{graphicx}
\usepackage{balance}
\usepackage{xcolor}
\usepackage{multirow}
\usepackage{makecell}
\usepackage{comment}
\usepackage{hyperref}
\usepackage{booktabs}
\usepackage{array}
\usepackage{lipsum} 

\newcolumntype{C}[1]{>{\centering\arraybackslash}m{#1}}
\newcolumntype{R}{>{\centering\arraybackslash}X}
\newcolumntype{L}{>{\raggedright\arraybackslash}X}

\begin{document}

\title{Spatial Characterization of Sub-Synchronous Oscillations Using Black-Box IBR Models}

\author{{Muhammad~Sharjeel~Javaid},~\IEEEmembership{Member,~IEEE}, {Gabriel~Covarrubias~Maureira},~\IEEEmembership{Student~Member,~IEEE}, {Ambuj~Gupta},~\IEEEmembership{Student~Member,~IEEE}, {Debraj~Bhattacharjee},~\IEEEmembership{Student~Member,~IEEE},\\ {Jianli~Gao},~\IEEEmembership{Member,~IEEE}, {Balarko~Chaudhuri},~\IEEEmembership{Fellow,~IEEE}, and {Mark~O'Malley},~\IEEEmembership{Fellow,~IEEE} 
    \thanks{This work has been supported by Engineering and Physical Sciences Research Council (EPSRC) under Grant EP/Y025946/1, and by the Leverhulme International Professorship under Grant LIP-2020-002. \textit{(Corresponding~author: Jianli Gao.)}}
}

\maketitle

\begin{abstract}
Power systems with high penetration of inverter-based resources (IBRs) are prone to sub-synchronous oscillations (SSO). The opaqueness of vendor-specific IBR models limits the ability to predict the severity and the spread of SSO. This paper demonstrates that black-box IBR models estimated through frequency-domain identification techniques, along with dynamic network model can replicate the actual oscillatory behavior. The estimated IBR models are validated against actual IBR models in a closed-loop multi-IBR test system through modal analysis by comparing closed-loop eigenvalues, and participation factors. Furthermore, using output-observable right eigenvectors, spatial heatmaps are developed to visualize the spread and severity of dominant SSO modes. The case studies on the 11-bus and 39-bus test systems confirm that even with the estimated IBR models, the regions susceptible to SSO can be identified in IBR-dominated power systems.
\end{abstract}

\begin{IEEEkeywords}
Inverter-based resources, observability, sub-synchronous oscillations, system identification.
\end{IEEEkeywords}

\section{Introduction}
Sub-synchronous oscillation (SSO) is a well-reported operational risk in grids with high penetrations of inverter-based resources (IBRs). Recent field reports mention multiple real-world SSO incidents across North America and Europe and show that weak grid conditions, tightly tuned inverter controls, and complex IBR–network interactions can trigger poorly damped SSO, leading to protection trips, large power swings, and repeated curtailment of renewable generation \cite{Cheng2023Real}. These events highlight that SSO in modern systems is no longer confined to classic turbine–generator torsional modes, but is increasingly driven by control-interaction modes of IBRs coupled through the network.

Replicating such events in simulation is essential for understanding root causes and designing mitigation measures, yet this task is complicated by the widespread use of vendor-specific, black-box IBR models. In many EMT and RMS studies the inverter dynamics are only accessible through proprietary blocks, which prevents the construction of an explicit state-space model for the full system and limits conventional modal analysis, participation factors, and spatial characterization of critical modes. By contrast, when a small-signal model of the entire IBR-dominated system is available, whether in state-space or an equivalent admittance form, eigenvalue sensitivity and participation analysis can identify which IBRs and network locations contribute most to each poorly damped mode \cite{Zhu2022Greybox, blackboxpal}. Building on this insight, we present how to obtain system-level dynamic representation using estimated IBR models which can be used to visualize SSO modes emergence and propagation through the network.



As its primary contribution, this work presents a unified workflow that links black-box IBR model estimation with system-level small-signal stability assessment and spatial visualization of SSO risk. Rather than proposing another identification algorithm, we provide an overview of three complementary estimation families: 1) frequency-sweep impedance scans, 2) wide-band signal injection methods, and 3) Eigensystem Realization Algorithm (ERA)-based time-domain identification. In this work, we employ wide-band frequency-domain perturbation followed by Fourier Transform and vector fitting to obtain estimated IBR models that reproduce the critical sub-synchronous modes, their modal composition, and spatial observability with sufficient fidelity and hence can be used to generate accurate heatmaps of SSO-susceptible regions. The other two estimation families are briefly discussed to contextualize the workflow and highlight alternative routes to obtain the estimated IBR models.


\section{IBR Model Estimation Techniques}
\label{sec:est_tech}

In this section, we briefly review the estimation techniques used to obtain IBR admittance models that feed into the system-wide modal analysis required for spatial characterization of SSO.

\begin{figure}[!t]
\centering
\includegraphics[width = 1\linewidth]{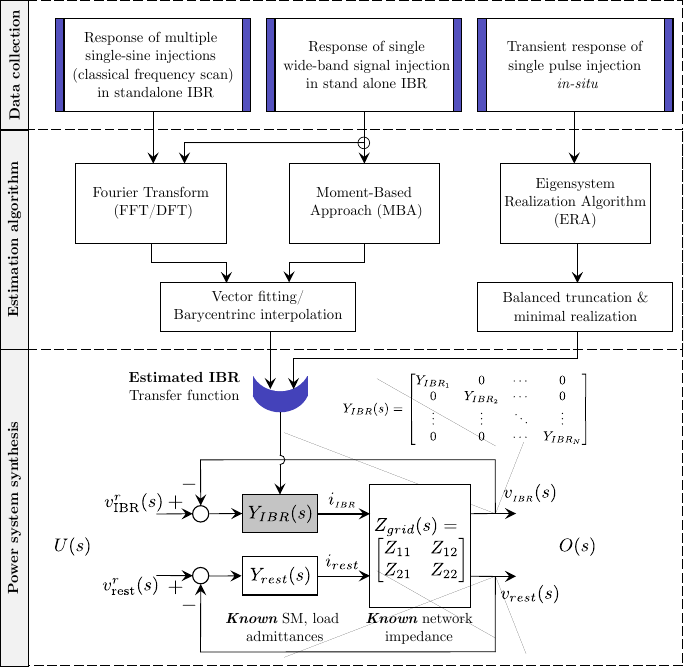}
\caption{Power system modeling with estimated IBR admittance transfer functions.}
\label{fig:flowchart1}
\end{figure}

As shown in Fig.~\ref{fig:flowchart1}, a key step in the workflow is to obtain a small-signal admittance model for every black-box IBR. In this paper we group existing techniques into three families:
(i) classical dynamic impedance scans based on multiple single-sine injections;
(ii) wide-band single-shot injections; and
(iii) time-domain identification using ERA. Each method eventually provides a rational approximation of the IBR admittance matrix $Y_{\text{IBR}}(s)$ in the synchronous $dq$ frame obtained using vector fitting or barycentric interpolation \cite{barycentric}, as illustrated in Fig.~\ref{fig:flowchart1}.

Dynamic impedance scan (frequency-sweep) is the de facto standard in many impedance-based stability studies and industrial tools. The system is perturbed with a sequence of low-amplitude sinusoidal signals at frequencies $\omega_k$ within a specific frequency band, and the magnitude and phase of the output signal is extracted via FFT/DFT to obtain discrete samples of the admittance $Y_{\text{IBR}}(j\omega_k)$. Depending on the implementation, the injection is applied either in the $dq$ control frame or in sequence components, and the resulting multi-input--multi-output (MIMO) admittance is assembled from several experiments. This methodology underpins recent impedance-ratio-based stability assessments for multi-inverter systems and dedicated frequency-domain tools such as INOSA and NREL's impedance-scan workflows, and is widely used to characterize grid-following/grid forming inverters (GFL/GFM) over a broad frequency range \cite{GIST2023Shah}.
Although mathematically simple and robust to noise, the need to wait for steady state at each frequency makes fine-resolution scans slow (hours to days for realistic EMT models), motivating faster alternatives.

To reduce measurement time, wide-band ``one-shot'' injection techniques excite the IBR with a rich spectrum (multi-sine, PRBS, or carefully designed wide-band waveforms) and recover the frequency response from a single experiment. Recent work on grid-impedance identification shows that such wide-band injections, combined with parametric or non-parametric processing, can reconstruct MIMO $dq$ impedances over tens to hundreds of Hz \cite{Gong2025Moment}.

In the third branch, ERA-based time-domain identification is considered for scenarios where only pulse or switching transients are available (e.g., in-situ disturbance records or staged tests). Starting from the sampled pulse response, ERA constructs a block Hankel matrix, performs an SVD-based rank reduction, and recovers a balanced minimal realization $(A,B,C,D)$ that captures the dominant oscillatory modes of the IBR \cite{Fan2021TDM}.
The resulting model can be rescaled to an equivalent admittance form $Y_{\text{IBR}}(s)$ which naturally incorporates damping and modal coupling.


Although all three branches can provide accurate models, some degree of mismatch with the underlying ``true'' inverter dynamics is inevitable owing to discrete frequency resolution, measurement time window, and model order selection/reduction. In Section \ref{sec:11_bus} we show that, even when the identified eigenvalues exhibit some differences, mode shapes and participation factors associated with SSO modes are preserved to a high degree, and the resulting heat-map visualizations remain similar to actual model simulations. Since network and conventional generation models are typically known to the system operator, the complete small-signal model is obtained by synthesizing the estimated IBR admittances with the known synchronous-machine, load, and network models, as depicted in Fig.~\ref{fig:flowchart1}. It is important to note that these estimation techniques are not exhaustive but represent the most common approaches available in the literature.

\section{Modal Analysis Overview}
\label{sec:full_system}
Considering the feedback system shown in the bottom of Fig.\ref{fig:flowchart1}, where $Y_{\text{rest}}(s)$ denotes the block-diagonal small-signal admittance of the all known dynamic models, i.e., synchronous generators and loads, and $v^{r}(s)$ denotes the reference voltage signal, the system transfer function can be derived as follows:

\begin{align}
-\underbrace{
\begin{bmatrix}
    i_{\text{IBR}}(s) \\
    i_{\text{rest}}(s)
\end{bmatrix}
}_{\mathbf{i}(s)}
&=
\underbrace{
\begin{bmatrix}
    Y_{IBR}(s) & 0\\
    0 & Y_{rest}(s)
\end{bmatrix}
}_{Y_{m}(s)}
\cdot
(
v^r(s)-\underbrace{\begin{bmatrix}
    v_{\text{IBR}}(s) \\
    v_{\text{rest}}(s)
\end{bmatrix}}
_{\mathbf{v}(s)})
\label{eq:eq-1}
\end{align}
Multiplying (\ref{eq:eq-1}) by known network impedance, $\mathbf{Z_{grid}}(s)$, on both sides yields:
\begin{align}
\mathbf{v}(s) = \mathbf{Z_{grid}}(s)\cdot \mathbf{Y_m(s)} \cdot (v^r(s)-\mathbf{v}(s)) 
\label{eq:eq-2}
\end{align}
Finally, by rearranging terms, the resulting transfer function, $\mathbf{G^{sys}}(s)$, is expressed as:
\begin{align}
\mathbf{G^{sys}}(s) = (\mathbb{I}+\mathbf{Z_{grid}}(s)\cdot \mathbf{Y_m}(s))^{-1} \cdot \mathbf{Z_{grid}}(s)\cdot \mathbf{Y_m}(s)
\label{eq:eq-3}
\end{align}
The whole-system representation $\mathbf{G^{sys}}(s)$ can be expressed through the state-space matrices $(A^{\mathbf{G}}, B^{\mathbf{G}}, C^{\mathbf{G}}, D^{\mathbf{G}})$. Next, we provide the expression that tells how these state-space matrices can be used to assess system-wide observability of SSO and IBR's participation in it. In this work, the system is modeled in the $dq$ reference frame, which has been shown to be suitable for small-signal analysis \cite{emt-rms}; therefore, each IBR and network component is modeled as a 2×2 MIMO system.

\subsection{System-wide Observability of Oscillation}
 The observability of a given mode $\lambda_i$ can be obtained as:
\begin{equation}
    \text{Obs}_{\lambda_i} = C^{\mathbf{G}}\cdot \psi_{:,i}
\end{equation}
where $\psi_{:,i}$ corresponds to the right eigenvector of $A^{\mathbf{G}}$ associated with the eigenvalue $\lambda_i$. Following a disturbance that excites a specific mode, it is expected that the buses with higher observability exhibit more sustained voltage oscillations (i.e., higher outputs of $\mathbf{G^{sys}}$). Since the output variables of the whole-system representation are $v_D$ and $v_Q$, the observability of the voltage magnitude can be expressed as:
\begin{equation}
\label{eq:obs}
    \text{Obs}^{|v|}_{\lambda_i} = \sqrt{(\text{Obs}^{v_D}_{\lambda_i})^2+(\text{Obs}^{v_Q}_{\lambda_i}})^2
\end{equation}

\subsection{IBR Participation Factor}

For a given mode $\lambda_i$, in addition to $\psi_i$, we also obtain $\phi_i$, the left eigenvectors of $A^{\mathbf{G}}$. We first extract the entries associated with the IBR states, denoted by $\psi_i^{\text{IBR}}$ and $\phi_i^{\text{IBR}}$. The cumulative participation magnitude of the IBR states
in mode $\lambda_i$ is then defined as:
\begin{equation}
    \mathcal{P}_{\lambda_i}^{\text{IBR}} = \bigl|\phi_i^{\text{IBR}} \odot \psi_i^{\text{IBR}}\bigr|,
\end{equation}
where $\odot$ denotes the element-wise (Hadamard) product. Finally, a normalized IBR participation index is obtained as:
\begin{equation}
    \tilde{\mathcal{P}}_{\lambda_i}^{\text{IBR}} = 
    \frac{\mathcal{P}_{\lambda_i}^{\text{IBR}}}{\max_j \mathcal{P}_{\lambda_i}^{\text{IBR}_j}}.
\end{equation}

This index indicates how strongly the aggregate IBR dynamics contribute to each mode relative to the most IBR-dominated mode in the spectrum.

\section{Validity of Estimation Technique}
\label{sec:11_bus}
\begin{figure}[!t]
    \centering
    \includegraphics[width = 1\linewidth]{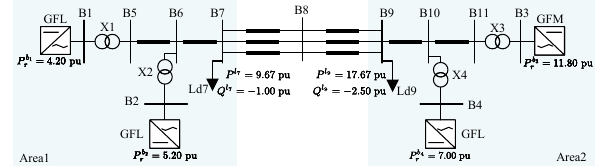}
    \caption{Kundur 11-bus test system with three GFLs and one GFM \cite{kundur1994power}.}
    \label{fig:kundur_c}
\end{figure}

\begin{figure}[!t]
    \centering
    \includegraphics[width = 1\linewidth]{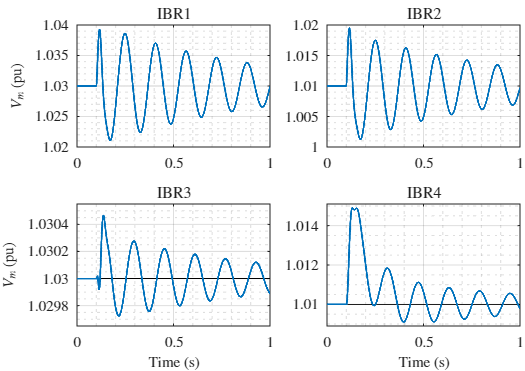}
    \caption{SSO observed in voltage magnitude of all four IBRs when power injection from IBR2 is increased by 5\% (from 5.2 pu to 5.46 pu).}
    \label{fig:TimeDomainSSO}
\end{figure}

To assess the validity of the proposed estimation approach, the modified 11-bus system shown in Fig. \ref{fig:kundur_c} is used as a test case. Three changes are made to the original network \cite{kundur1994power}. First, all synchronous generators are replaced by IBRs: buses 1, 2, and 4 host grid-following (GFL) units, while bus 3 hosts a grid-forming (GFM) unit (this information is not required by the estimation algorithm itself). Second, the transmission corridor between buses 7 and 9 is reinforced by adding a third parallel line, resulting in three lines between these buses. Third, the active power injections at buses 1 and 2 are set to 4.2 pu and 5.2 pu. Under this operating condition, the system exhibits a poorly damped SSO mode at approximately 6.4 Hz. This is illustrated in Fig. \ref{fig:TimeDomainSSO}, which shows the time-domain response of the voltage magnitudes at the four IBR buses when the power injection from IBR2 is increased from 5.2 pu to 5.46 pu (a 5\% step). IBR1 and IBR2, located in Area1, experience large-amplitude oscillations, while IBR3 and IBR4 in Area 2 show noticeably smaller swings. This behavior indicates that the dominant SSO mode is mainly localized in Area1, with weaker coupling to Area 2. Now, the objective is to confirm that this can be assessed through modal analysis of a system where estimated IBR models are used. 

\begin{figure}[!t]
    \centering
    \includegraphics[width = 1\linewidth]{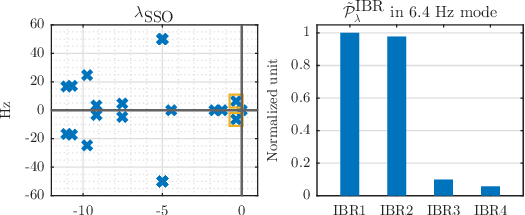}
    \caption{Sub-synchronous eigenvalues $\lambda_\textrm{SSO}$ of the model-based closed-loop 11-bus test system (left). Normalized participation of IBRs $\tilde{\mathcal{P}}_\textrm{IBR}$ in the least damped SSO $\lambda_{\textrm{SSO}_1}$ (right).}
    \label{fig:ActualEV_PartFact}
\end{figure}

The outcome of closed-loop system modal analysis using the actual detailed IBR models serves as the benchmark. The resulting eigenvalues of the linearized system and the corresponding normalized participation factors for the least damped SSO mode are shown in Fig. \ref{fig:ActualEV_PartFact}. The mode around 6.4 Hz is highlighted with yellow marker on the left, and its participation profile indicates that IBR1 and IBR2 are the primary contributors, as expected.

\begin{figure}[!t]
    \centering
    \includegraphics[width = 1\linewidth]{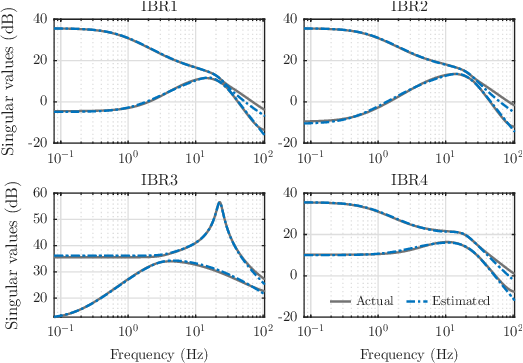}
    \caption{Comparison of singular values of estimated admittance transfer function with the singular values of actual model admittance of all four IBRs.}
    \label{fig:SigmaComparison}
\end{figure}

\begin{figure}[!t]
    \centering
    \includegraphics[width = 1\linewidth]{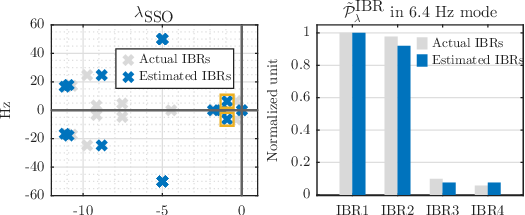}
    \caption{Comparison of sub-synchronous eigenvalues $\lambda_\textrm{SSO}$ (left) and normalized participation in the least-damped mode $\tilde{\mathcal{P}}_\textrm{IBR}$ (right) of estimated IBR-based closed-loop 11-bus system with those obtained using actual IBR model.}
    \label{fig:ComparisonEV_PF}
\end{figure}

For estimation, each IBR is placed in a standalone setup, connected to an ideal voltage source that enforces the same steady-state terminal conditions as in the network. A single wide-band perturbation, with frequency content spanning 0.1–100 Hz, is injected from the voltage source. In the time domain, the resulting $dq$-frame currents at the IBR terminals are measured. Applying the FFT to both the injected voltage and the measured current, and computing their ratio, yields the complex admittance spectra. For each IBR, 2×2 MIMO admittance transfer function ($dq$ voltages as inputs, $dq$ currents as outputs) is then obtained using vector fitting. The sigma plots of the estimated IBR models are compared against those obtained from the actual IBR models in Fig. \ref{fig:SigmaComparison} and show good agreement over the targeted frequency range (0.1~Hz-100~Hz).
Finally, the closed-loop system is synthesized by interconnecting the linearized network model with the estimated IBR admittance transfer functions. The sub-synchronous eigenvalues and associated participation factors, shown in Fig. \ref{fig:ComparisonEV_PF}, closely match the benchmark results. 

\begin{figure}[!t]
    \centering
    \includegraphics[width = 1\linewidth]{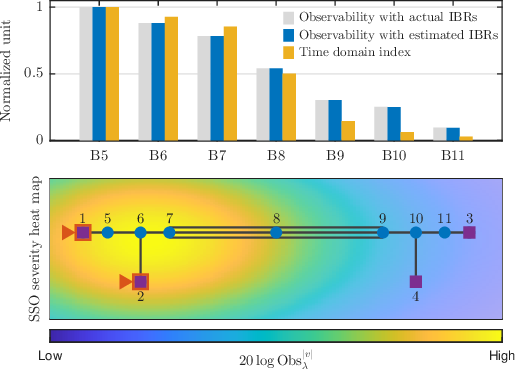}
    \caption{Top: Comparison of voltage magnitude observability indices. Bottom: SSO severity and its widespread observability. Blue circles are non-IBR buses, purple squares are IBR buses and red arrows point to the IBRs with the highest participation in 6.4 Hz mode.}
    \label{fig:KundurBars_HeatMaps}
\end{figure}

In Fig. \ref{fig:KundurBars_HeatMaps} SSO severity obtained from modal analysis using actual and estimated IBR models is compared with time-domain observations and the subsequent heatmap is presented. The upper subplot shows the normalized observability index of the critical 6.4 Hz mode at non-IBR buses B5–B11, computed using \eqref{eq:obs} with the actual IBR models (grey) and the estimated admittance models (blue), alongside a time-domain severity index derived from the maximum oscillation amplitudes at these buses (yellow). Both sets of modal observability indices closely follow the ranking implied by the time-domain index, indicating that the estimated models correctly identify the buses most affected by the mode. The lower subplot presents a spatial heatmap of the same observability measure (in dB), where blue circles denote non-IBR buses, purple squares denote IBR buses, and arrow markers highlight the IBRs that predominantly contribute to the mode. This visually confirms the concentration of dominant SSO activity in Area1.

\section{Application on 39-Bus Test System}
\label{sec:result_sec}

\begin{figure}[!t]
    \centering
    \includegraphics[width=0.95\linewidth]{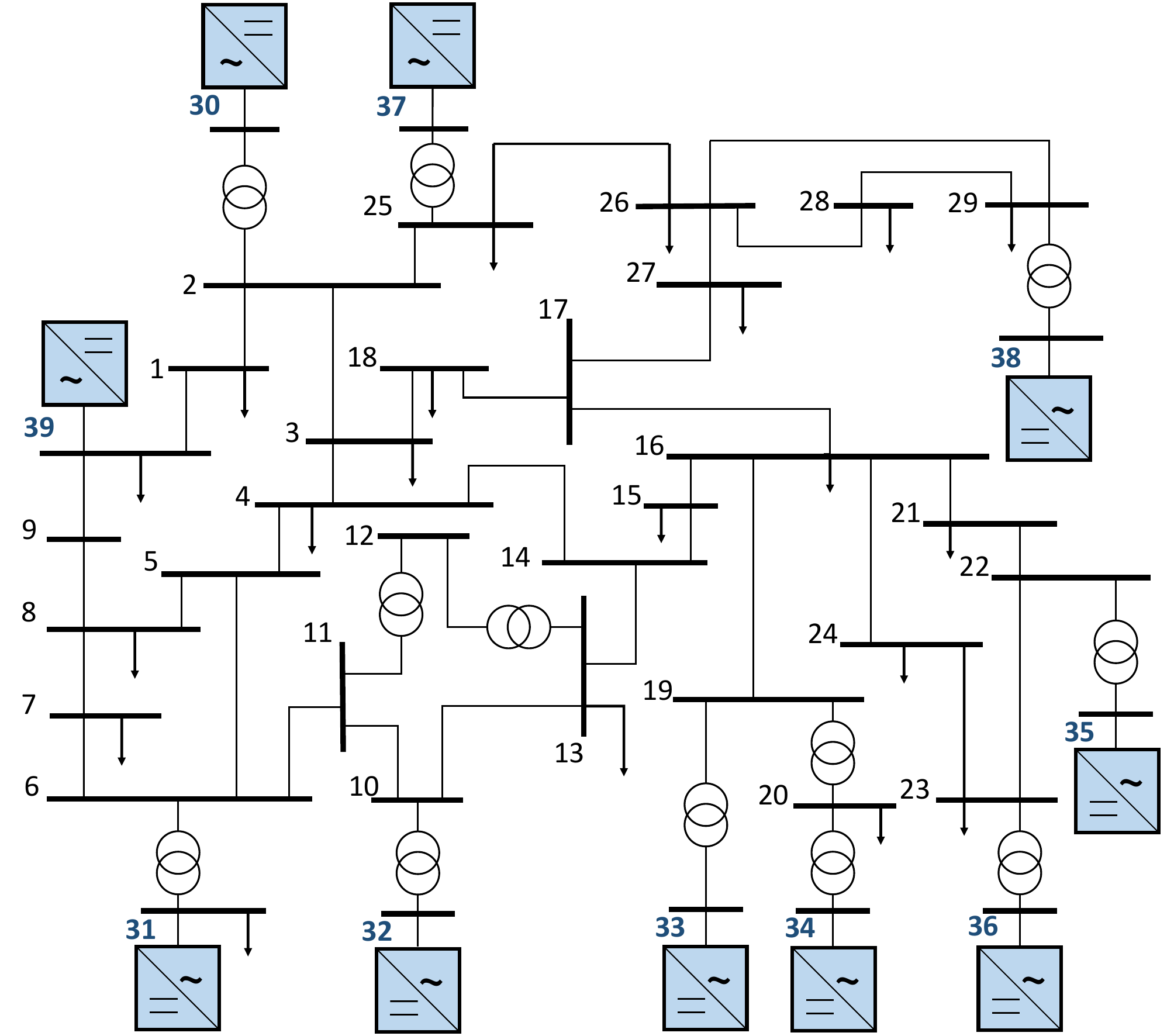}
    \caption{IEEE 39-bus power system \cite{ref39bus}, where all machines are replaced by IBRs. See Table \ref{tab:gfl_gfm_scenarios} for the type and location of each IBR in different scenarios.}
    \label{fig:39Bus_c}
\end{figure}

The proposed framework is next applied to the IEEE 39-bus test system \cite{ref39bus}, in which ten synchronous generators are replaced by IBRs at buses 30–39, as shown in Fig. \ref{fig:39Bus_c}. Four operating scenarios are defined by relocating GFL and GFM units and adjusting their active power injections, as summarized in Table \ref{tab:gfl_gfm_scenarios}, so that both the spatial placement of IBRs and the GFL/GFM mix vary across scenarios.

\begin{table}[ht!]
\centering
\caption{IBR Configurations for the Four Scenarios}
\label{tab:gfl_gfm_scenarios}
\renewcommand{\arraystretch}{1.25}
\setlength{\tabcolsep}{4pt}
\footnotesize
    \begin{tabular}{|c|c|c|c|c|c|c|c|c|}
    \hline
    \textbf{Bus} & \multicolumn{2}{c|}{\textbf{Scenario~I}} & \multicolumn{2}{c|}{\textbf{Scenario~II}} & \multicolumn{2}{c|}{\textbf{Scenario~III}} & \multicolumn{2}{c|}{\textbf{Scenario~IV}} \\
    \cline{2-9}
    & Type & $P$ (pu) & Type & $P$ (pu) & Type & $P$ (pu) & Type & $P$ (pu) \\
    \hline
    30 & GFL & 3.55  & GFM & 2.75 & GFL & 2.5 & GFM & 2.5 \\
    31 & GFM & -5.86 & GFL & 4.99 & GFL & 8.76 & GFM & 12.35 \\
    32 & GFL & 9.23  & GFM & 6.5  & GFL & 6.5 & GFM & 6.5 \\
    33 & GFM & 6.32  & GFL & 6.32 & GFL & 6.32 & GFM & 6.32 \\
    34 & GFL & 7.21  & GFM & 5.08 & GFL & 5.08 & GFM & 5.08 \\
    35 & GFM & 6.5   & GFL & 6.5   & GFM & 5.85 & GFL & 5.2 \\
    36 & GFL & 7.95  & GFM & 5.6  & GFM & 5.04 & GFL & 4.48 \\
    37 & GFM & 5.4   & GFL & 5.4  & GFL & 4.86 & GFL & 4.32 \\
    38 & GFL & 11.79 & GFM & 8.3  & GFM & 7.47 & GFL & 6.64 \\
    39 & GFM & 10    & GFL & 10 & GFM & 9 & GFL & 8 \\
    \hline
    \multicolumn{1}{|l|}{GFL} & \multicolumn{2}{c|}{54.1\%} & \multicolumn{2}{c|}{64\%} & \multicolumn{2}{c|}{47.5\%} & \multicolumn{2}{c|}{46.7\%} \\
    \multicolumn{1}{|l|}{GFM} & \multicolumn{2}{c|}{45.9\%} & \multicolumn{2}{c|}{36\%} & \multicolumn{2}{c|}{52.5\%} & \multicolumn{2}{c|}{53.3\%} \\
    \hline
    \end{tabular}
\end{table}

\begin{figure*}[!t]
    \centering
    \includegraphics[width = 1\linewidth]{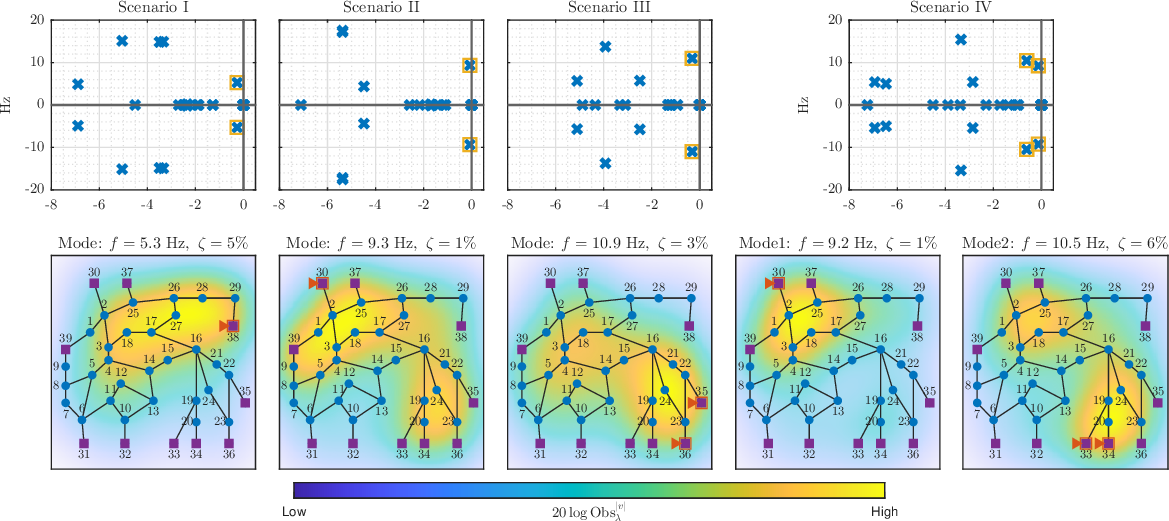}
    \caption{Top: Eigenvalues for each scenario, with undamped SSO modes marked. Bottom: Wide-spread observability and severity of SSO modes for each scenario. The IBRs with the highest participation factors are indicated by red arrows.}
    \label{fig:39Bus_HeatMaps}
\end{figure*}

 For each scenario, the range of sub-synchronous eigenvalues and the corresponding SSO severity heatmaps are shown in Fig. \ref{fig:39Bus_HeatMaps}. In Scenario I, the least-damped mode at 5.3 Hz is primarily driven by the IBR at bus 38, and the associated oscillations remain largely confined to the upper part of the network. In Scenario II, the IBR at bus 30 has the highest participation in the 9.3 Hz mode, yet the largest observability is at buses 2 and 3, with weaker traces around buses 23–24. This suggests that the bus(es) contributing most to a mode does not necessarily coincide with the buses where it is most visible. Scenario III features a 10.9 Hz mode with two dominant IBR contributors; the mode is strongly observable on nearby buses but still leaves a footprint in more remote parts of the grid. Finally, Scenario IV exhibits two poorly damped modes: a 9.2 Hz mode largely confined to the vicinity of buses 30 and 37, and a 10.5 Hz mode that simultaneously excites two separate regions of the network.
 
 Together, these cases demonstrate that the proposed modal-heatmap framework based on estimated IBR models scales to larger systems and can distinguish between relatively localized and system-wide SSO patterns under different IBR deployment strategies. The resulting heatmaps remain largely valid if the estimation errors in the IBR models do not move the poorly-damped mode in the closed-loop system beyond certain modal frequency and damping thresholds, nor alter the ranking of observability/participation used for spatial visualization. This robustness can be assessed via eigenvalue-sensitivity analysis which is identified as future work.

\section{Conclusion}
This paper has demonstrated that accurately estimated $dq$-frame admittance models of IBRs can be combined with a dynamic network model to reproduce the key sub-synchronous oscillation characteristics of IBR-dominated systems. On the modified 11-bus test system, the estimated models yielded sub-synchronous eigenvalues, participation factors, and bus-level observability indices that closely matched those obtained from the actual IBR models, and the resulting heatmaps correctly identified the area where the poorly-damped mode is most severe. Application to the IEEE 39-bus system showed that the proposed modal–heatmap framework scales to larger networks. The results underline that the buses contributing most to the SSO mode do not necessarily coincide with the buses where it is most observable, an important insight for designing monitoring and protection schemes. Overall, this paper provides a practical basis that can be extended to develop SSO early-warning tools for future control rooms even when detailed vendor-specific IBR models are unavailable.

\printbibliography

\end{document}